\documentclass[numberedappendix,12pt,preprint]{emulateapj}  

\usepackage[]{amsmath}
\usepackage{graphicx}
\usepackage{multirow}
\usepackage{booktabs}
\usepackage{bm}
\usepackage{xspace}
\usepackage{afterpage}
\usepackage{url}
\usepackage{textcomp}

\usepackage{color}
\usepackage[usenames,dvipsnames]{xcolor}
\definecolor{midgray}{gray}{0.4}
\usepackage[colorlinks=true,citecolor=blue,linkcolor=magenta]{hyperref}

\newcommand{\HST}{\textit{HST}\xspace}
\newcommand{\spitzer}{\textit{Spitzer}\xspace}
\newcommand{\photz}{z_{\mathrm{phot}}}

\newcommand{\mubest}{\mu_{\mathrm{best}}}
\newcommand{\msolar}{M_{\odot}\xspace}
\newcommand{\sfrunit}{\msolar\,\mathrm{yr}^{-1}\xspace}
\newcommand{\ssfrunit}{\mathrm{Gyr}^{-1}\xspace}

\newcommand{\lya}{Ly$\alpha$\xspace}
\newcommand{\civ}{C \textsc{IV}\xspace}
\newcommand{\ciii}{C \textsc{III}]\xspace}
\newcommand{\oii}{[O \textsc{II}]\xspace}
\newcommand{\oiii}{[O \textsc{III}]\xspace}
\newcommand{\luv}{L_{\mathrm{UV}}}
\newcommand{\degree}{$^{\circ}$\xspace}
\newcommand{\pasa}{Publications of the Astron. Soc. of Australia\xspace}

\begin{document}
\title{Detection of Lyman-Alpha Emission from a Triply-Imaged \lowercase{$z=6.85$} Galaxy Behind MACS J2129.4$-$0741}

\author{Kuang-Han Huang\altaffilmark{1}}
\author{Brian C. Lemaux\altaffilmark{1}}
\author{Kasper B. Schmidt\altaffilmark{2}}
\author{Austin Hoag\altaffilmark{1}}
\author{Maru\v{s}a Brada\v{c}\altaffilmark{1}}
\author{Tommaso Treu\altaffilmark{3}}
\author{Mark Dijkstra\altaffilmark{4}}
\author{Adriano Fontana\altaffilmark{5}}
\author{Alaina Henry\altaffilmark{6}}
\author{Matthew Malkan\altaffilmark{3}}
\author{Charlotte Mason\altaffilmark{3}}
\author{Takahiro Morishita\altaffilmark{3,7}}
\author{Laura Pentericci\altaffilmark{5}}
\author{Russell E. Ryan, Jr.\altaffilmark{8}}
\author{Michele Trenti\altaffilmark{9}}
\author{Xin Wang\altaffilmark{3}}

\altaffiltext{1}{University of California Davis, 1 Shields Avenue, Davis, CA 95616, USA; khhuang@ucdavis.edu}
\altaffiltext{2}{Leibniz-Institut f\"{u}r Astrophysik Potsdam (AIP), An der Sternwarte 16, 14482, Potsdam, Germany}
\altaffiltext{3}{Department of Physics and Astronomy, UCLA, Los Angeles, CA 90095, USA}
\altaffiltext{4}{Institute of Theoretical Astrophysics, University of Oslo, P.O. Box 1029, NO-0315 Oslo, Norway}
\altaffiltext{5}{INAF Osservatorio Astronomico di Roma, Via Frascati 33,00040 Monteporzio (RM), Italy}
\altaffiltext{6}{Astrophysics Science Division, Goddard Space Flight Center, Code 665, Greenbelt, MD 20771, USA}
\altaffiltext{7}{Astronomical Institute, Tohoku University, Aramaki, Aoba, Sendai 980-8578, Japan}
\altaffiltext{8}{Space Telescope Science Institute, Baltimore, MD 21218, USA}
\altaffiltext{9}{School of Physics, The University of Melbourne, VIC 3010, Australia}

\email[E-mail:~]{khhuang@ucdavis.edu}
\begin{abstract}

We report the detection of Ly$\alpha$ emission at $\sim9538$\AA{} in the Keck/DEIMOS and \HST WFC3 G102 grism data from a triply-imaged galaxy at $z=6.846\pm0.001$ behind galaxy cluster MACS J2129.4$-$0741. Combining the emission line wavelength with broadband photometry, line ratio upper limits, and lens modeling, we rule out the scenario that this emission line is \oii at $z=1.57$. After accounting for magnification, we calculate the weighted average of the intrinsic \lya luminosity to be $\sim1.3\times10^{42}~\mathrm{erg}~\mathrm{s}^{-1}$ and \lya equivalent width to be $74\pm15$\AA{}. Its intrinsic UV absolute magnitude at 1600\AA{} is $-18.6\pm0.2$ mag and stellar mass $(1.5\pm0.3)\times10^{7}~\msolar$, making it one of the faintest (intrinsic $\luv\sim0.14~\luv^*$) galaxies with \lya detection at $z\sim7$ to date. Its stellar mass is in the typical range for the galaxies thought to dominate the reionization photon budget at $z\gtrsim7$; the inferred \lya escape fraction is high ($\gtrsim 10$\%), which could be common for sub-$L^*$ $z\gtrsim7$ galaxies with \lya emission. This galaxy offers a glimpse of the galaxy population that is thought to drive reionization, and it shows that gravitational lensing is an important avenue to probe the sub-$L^*$ galaxy population.

\end{abstract}

\section{Introduction}

Galaxy cluster fields have become popular survey fields for the high-redshift universe, because strong gravitational lensing boosts the number counts in the bright end of the UV luminosity function (LF) and probes fainter intrinsic luminosities than in blank fields \citep[e.g.,][]{Coe:2015cx}. Intrinsically fainter galaxies at $z\gtrsim3$ are also more likely to exihibit \lya in emission, a result of correlation between UV luminosity and dust attenuation \citep[e.g.,][]{Schaerer:2011fr}. To constrain the properties of background galaxies, one needs a precise lensing map of the galaxy cluster, constructed using the positions and redshifts of multiply-imaged background galaxies \citep[e.g.,][]{Bradac:2009bk}; in particular, when a background galaxy has multiple images, their positions can be used to constrain the galaxy's redshift.

Several multiply-imaged $z\geq6$ galaxies have been discovered previously. Some have their redshifts confirmed by spectroscopy through the detection of their \lya emission \citep[e.g., ][]{Richard:2011il,Balestra:2013bj,Vanzella:2014cm}, while others are not confirmed by spectroscopy but have strong constraints from both photometry and lensing to be at redshifts up to $\sim11$ \citep[e.g.,][and references therein]{Zitrin:2014ew}. For the latter group of objects, gravitational lensing gives credence to the high-redshift interpretation even when no spectral features are detected, something that is not available in blank fields.

Here we report the detection of \lya emission, by both Keck/DEIMOS and \HST WFC3/IR grism, from three sources lensed by the galaxy cluster MACS J2129.4$-$0741 \citep[hereafter MACS2129;][]{Ebeling:2007fq} --- the highest-redshift multiply-imaged system spectroscopically confirmed to date. Two of the three sources (Images A and B) are selected as Lyman Break Galaxies (LBGs) at $z\sim 6-7$ by \citet{Bradley:2014gk}, and all three sources are considered multiple images of the same galaxy at $z=6.5$ in \citet{Zitrin:2015gu}. Based on their photometric, spectroscopic, and lensing constraints, the most natural explanation is that they are multiple images of the same galaxy at $z=6.85$. We discuss the photometry in Section \ref{sec:photom}, the spectroscopy in Section \ref{sec:spec}, the lens modeling in Section \ref{sec:lensing}, and their physical properties and implications for reionization in Section \ref{sec:discussion}. We adopt the cosmological parameters $H_0=70~\mathrm{km}~\mathrm{s}^{-1}~\mathrm{Mpc}^{-1}$, $\Omega_m=0.3$, and $\Omega_\Lambda=0.7$ in our analyses, and all magnitudes are in the AB system.

\begin{figure}[t]
\includegraphics[width=\columnwidth,trim={0 0cm 0 0},clip]{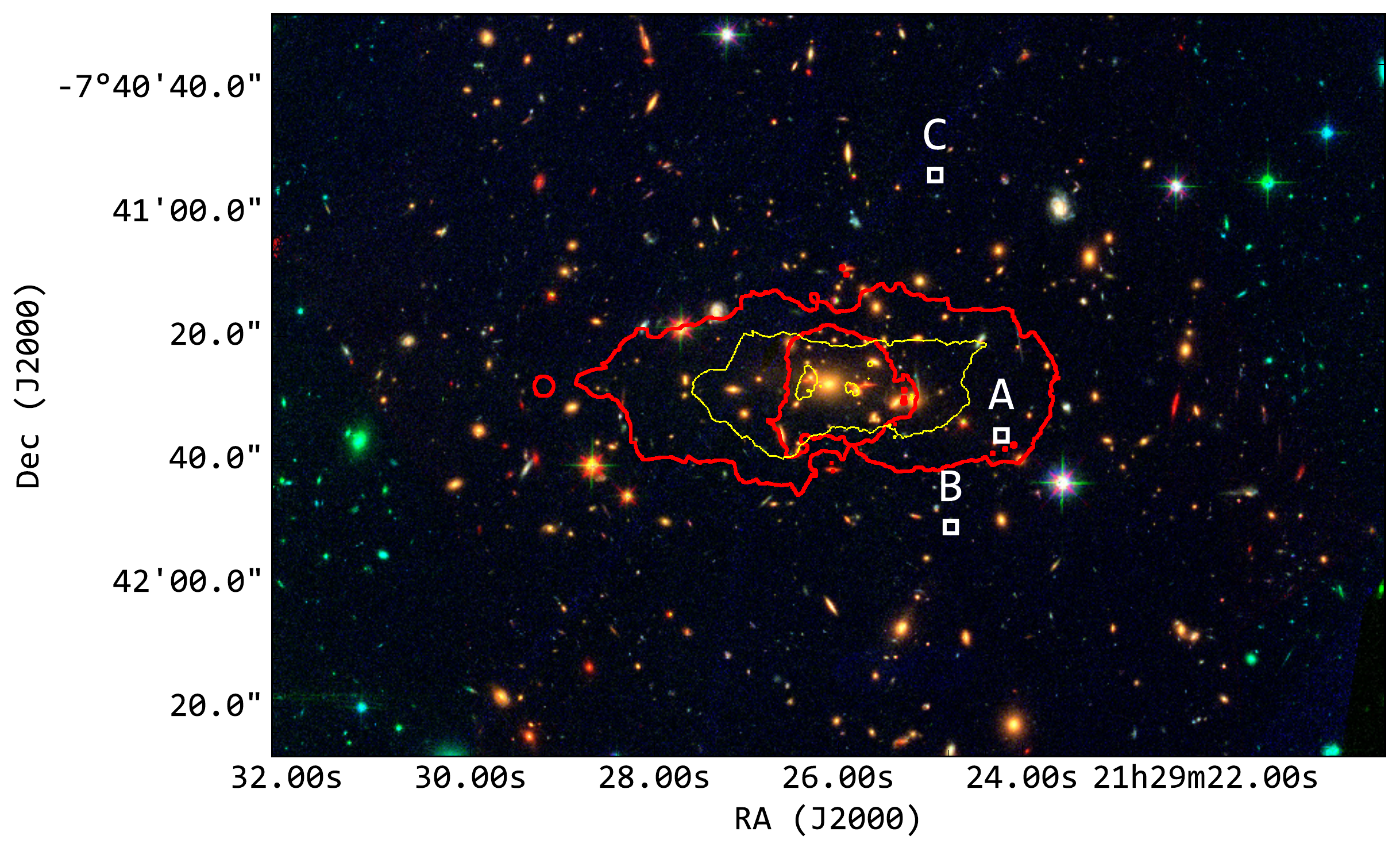}
\includegraphics[width=\columnwidth,trim={0 0cm 3cm 1cm},clip]{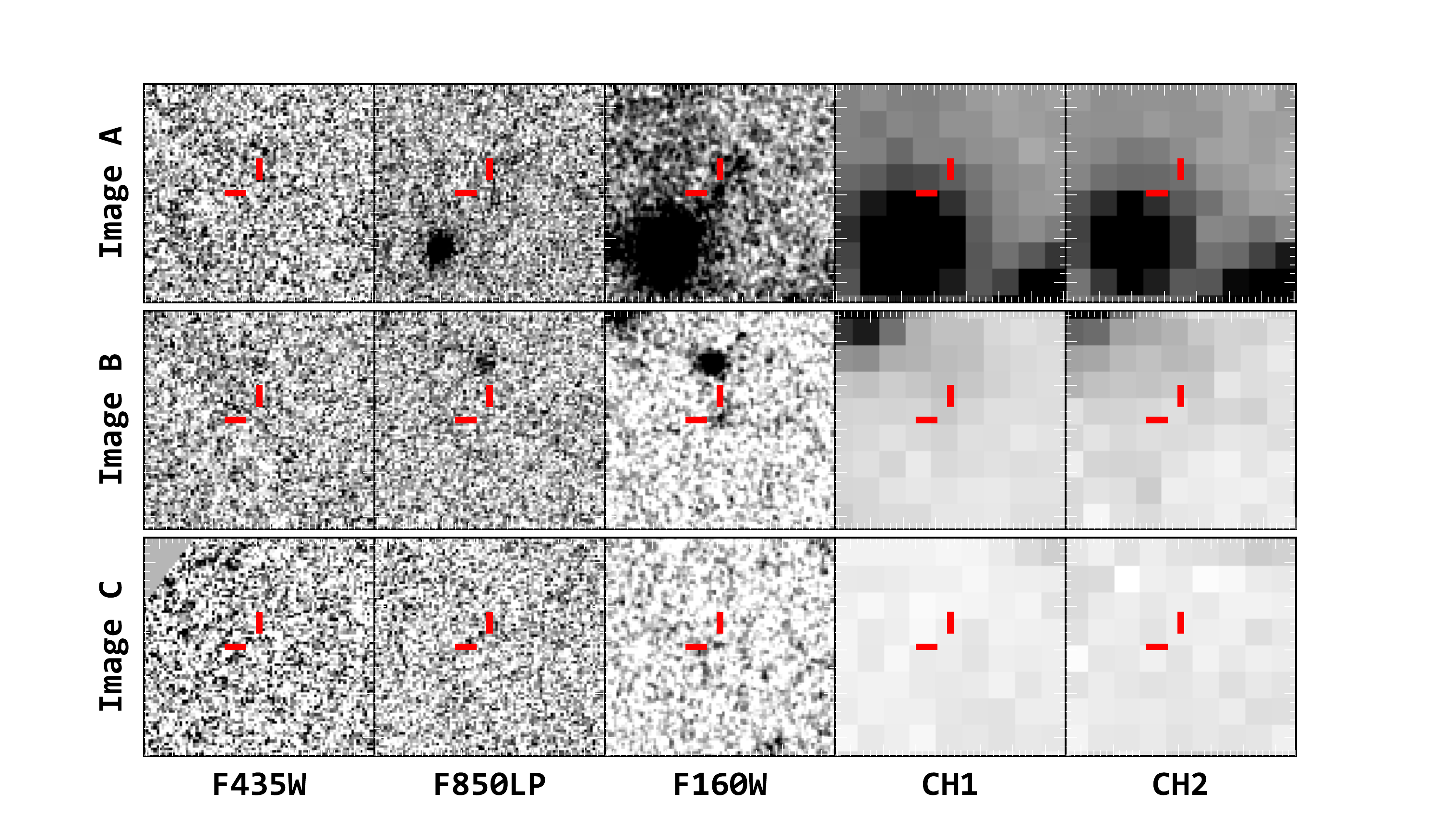}
\caption{\emph{Top:} RGB image covering the central $3\arcmin\times2\arcmin$ of MACS2129.4$-$0741 and the locations of Images A, B, and C. The critical curve ($\mu=100$) of the cluster at $z=6.85$ ($z=1.57$) is shown as red (yellow) lines. The $z=1.57$ critical curve does not come close to reproducing the parity of the triple images. \emph{Bottom:} Cutouts ($5\arcsec$ on each side) of all three images in F435W, F850LP, F160W, IRAC ch1, and IRAC ch2 (from left to right) are shown in each row. There is a nominal $3.3\sigma$ detection in the F850LP data for Image A that we consider spurious based on its morphology, and we exclude the F850LP magnitude of Image A in our analysis.\label{fig:macs2129}}
\end{figure}

\begin{deluxetable*}{llccc}
\tabletypesize{\scriptsize}
\tablecaption{Summary of Multiple Images\tablenotemark{a}\label{tab:summary}}

\tablecolumns{5}
\tablehead{\colhead{} & \colhead{} & \colhead{Image A} & \colhead{Image B} & \colhead{Image C}}
\startdata
R.A. & (deg.) & 322.350936 & 322.353239 & 322.353943 \\
Decl. & (deg.) & -7.693322 & -7.697442 & -7.681646 \\
\rule[-1ex]{0pt}{0pt}$\mubest$ &  & $11.0^{+0.1}_{-0.7}$ & $5.4^{+0.1}_{-0.1}$ & $2.6^{+0.1}_{-0.1}$ \\
\hline
\rule{0pt}{2ex}F225W & (mag) & $>26.30$ & $>27.07$ & $>27.77$ \\
F275W & (mag) & $>26.57$ & $>27.16$ & $>27.90$ \\
F336W & (mag) & $>26.75$ & $>27.51$ & $>28.21$ \\
F390W & (mag) & \nodata  & $>28.00$ & $>28.65$ \\
F435W & (mag) & $26.70 \pm 0.63$ & $>27.81$ & $>28.27$ \\
F475W & (mag) & $>27.53$ & $>28.17$ & $>28.79$ \\
F555W & (mag) & $27.64 \pm 0.71$ & $>28.58$ & $>29.25$ \\
F606W & (mag) & $> 27.76$ & $>28.39$ & $>28.73$ \\
F625W & (mag) & $26.64 \pm 0.69$ & $>27.99$ & $28.46 \pm 0.89$ \\
F775W & (mag) & $> 27.23$ & $27.89$ & $>28.48$ \\
F814W & (mag) & $27.07 \pm 0.52$ & $28.76 \pm 0.99$ & $>29.45$ \\
F850LP & (mag) & ($25.69 \pm 0.35$)\tablenotemark{b} & $>27.58$ & $>28.08$ \\
F105W & (mag) & $25.77 \pm 0.30$ & $26.33 \pm 0.21$ & $27.58 \pm 0.34$ \\
F110W & (mag) & $25.33 \pm 0.17$ & $26.42 \pm 0.23$ & $27.84 \pm 0.41$ \\
F125W & (mag) & $25.70 \pm 0.20$ & $26.42 \pm 0.21$ & $27.42 \pm 0.28$ \\
F140W & (mag) & $25.62 \pm 0.17$ & $26.63 \pm 0.24$ & $28.26 \pm 0.47$ \\
F160W & (mag) & $25.83 \pm 0.23$ & $26.64 \pm 0.21$ & $28.13 \pm 0.54$ \\
$[3.6]$ & (mag) & $>24.15$ & $26.19 \pm 0.51$ & $>27.12$ \\
\rule[-1ex]{0pt}{0pt}$[4.5]$ & (mag) & $>25.22$ & $>26.57$ & $>27.09$ \\
\hline
\rule{0pt}{2.5ex}$M_\mathrm{UV}-2.5\log(\mu/\mu_{\mathrm{best}})$ & (mag) & $-18.6 \pm 0.2$ & $-18.6 \pm 0.2$ & $-18.4 \pm 0.3$ \\
$M_*\times \mu/\mubest$ & ($10^7~\msolar$) & $1.7^{+0.6}_{-0.2}$ & $1.6^{+1.0}_{-0.04}$ & $1.3^{+0.3}_{-0.4}$ \\
$\mathrm{SFR}\times \mu/\mubest$ & ($\sfrunit$) & $1.8^{+0.3}_{-0.3}$ & $1.7^{+0.9}_{-0.3}$ & $0.8^{+0.2}_{-0.1}$ \\
sSFR & (Gyr$^{-1}$) & $105.1^{+0.0}_{-42.1}$ & $105.1^{+0.0}_{-42.1}$ & $65.7^{+39.4}_{-10.3}$ \\
Age & (Myr) & $\leq 17$ & $\leq 17$ & $17^{+3}_{-7}$ \\
$E(B-V)$\tablenotemark{c} & (mag) & $<0.05$ & $<0.05$ & $<0.05$ \\
$R_e\times\sqrt{\mu/\mu_{\mathrm{best}}}$\tablenotemark{d} & (kpc) & $0.8 \pm 0.3$ & $0.5 \pm 0.1$ & $0.4 \pm 0.1$ \\
$\beta$\tablenotemark{e} &   & $-2.49^{+1.39}_{-1.43}$ & $-2.96^{+1.35}_{-1.68}$ & $\leq -1.89$ \\
$\chi^2_{z=6.85}$ &  & 6.5 & 5.1 & 1.4 \\
\rule[-1.5ex]{0pt}{0pt}$\chi^2_{z=1.57}$ &  & 9.4 & 22.1 & 5.0  \\

\hline
\multicolumn{5}{c}{Keck DEIMOS Measurement}\\
$t_{\mathrm{exp}}$  &  (ks)  &  21.6  &  40.2  &  40.2 \\
\rule{0pt}{2.5ex}$f_{\mathrm{Ly}\alpha}^{\mathrm{DEIMOS}}$ & ($10^{-17}$ erg\,s$^{-1}$\,cm$^{-2}$) & $2.0 \pm 0.1$ & $1.1 \pm 0.1$ & $0.7 \pm 0.1$ \\
\rule{0pt}{2.5ex}$L_{\mathrm{Ly}\alpha}^{\mathrm{DEIMOS}}\times \mu/\mubest$ & ($10^{42}$ erg\,s$^{-1}$) & $1.0 \pm 0.1$ & $1.2 \pm 0.1$ & $2.0 \pm 0.3$ \\
\rule{0pt}{2.5ex}$W_{\mathrm{Ly}\alpha}^{\mathrm{DEIMOS}}$ & (\AA) & $60 \pm 11$ & $47 \pm 9$ & $170 \pm 77$ \\
\rule{0pt}{2.5ex}SFR$_{\mathrm{Ly}\alpha}^{\mathrm{DEIMOS}}\times \mu/\mubest$ & ($\sfrunit$) & $1.0 \pm 0.1$ & $1.1 \pm 0.1$ & $1.9 \pm 0.3$ \\
\hline
\multicolumn{5}{c}{\HST Grism Measurement}\\
\rule{0pt}{2.5ex}$f_{\mathrm{Ly}\alpha}^{\mathrm{PA328}}$ & ($10^{-17}$ erg\,s$^{-1}$\,cm$^{-2}$) & $3.5 \pm 0.8$ & $2.7 \pm 0.8$ & $2.1 \pm 0.8$ \\
\rule{0pt}{2.5ex}$f_{\mathrm{Ly}\alpha}^{\mathrm{PA050}}$ & ($10^{-17}$ erg\,s$^{-1}$\,cm$^{-2}$) & $1.7 \pm 1.0$ & $4.4 \pm 1.0$ & $2.6 \pm 1.0$ \\
\rule{0pt}{2.5ex}$W_{\mathrm{Ly}\alpha}^{\mathrm{PA328}}$ & (\AA) & $93 \pm 27$ & $100 \pm 33$ & $399 \pm 237$ \\
\rule{0pt}{2.5ex}$W_{\mathrm{Ly}\alpha}^{\mathrm{PA050}}$ & (\AA) & $43 \pm 27$ & $160 \pm 45$ & $513 \pm 300$ \\
\rule{0pt}{2.5ex}$W_{\mathrm{\civ}}^{\mathrm{PA328}}$ & (\AA) & $\leq 24$ & $\leq 39$ & $\leq 105$ \\
\rule{0pt}{2.5ex}$W_{\mathrm{\ciii}}^{\mathrm{PA050}}$ & (\AA) & $\leq 27$ & $\leq 43$ & $\leq 115$ 

\enddata

\tablenotetext{a}{All error bars and limits are $1\sigma$ values, and the errors for magnification factors $\mu_{\mathrm{best}}$ are random errors only.}
\tablenotetext{b}{The F850LP data for Image A has some spurious flux that led to a nominal $3.3\sigma$ detection, and we exclude it from our SED fitting procedure, although including it does not change our results significantly.}
\tablenotetext{c}{All three images have best-fit $E(B-V)=0$.}
\tablenotetext{d}{Effective radii are measured in the stacked WFC3/IR images.}
\tablenotetext{e}{The UV slope measured from the median-stacked \HST images is $-2.76^{+1.23}_{-1.12}$.}

\end{deluxetable*}

\section{Photometric Constraints}\label{sec:photom}

We use deep \HST and \spitzer imaging data for MACS2129 to derive photometric constraints for the three images. The \HST imaging data were taken as a part of the Cluster Lensing And Supernova survey with Hubble \citep[CLASH;][]{Postman:2012ca} program. 
We perform photometry in the same way as \citet{Huang:2016}: source detection is done in the coadded CLASH WFC3/IR image using SExtractor, and colors are measured in isophotal apertures. We do not match the PSFs of different \HST bands because 
convolving each band with a PSF-matching kernel introduces additional noise and degrades the signal-to-noise ratios. We run a simple check of our \HST photometry by fitting photometric redshifts ($\photz$) to all sources with S/N$\geq3$ in at least three filters, and confirm that the $\photz$ distribution peaks within $\Delta~z=\pm0.1$ of the cluster redshift, $z_{\mathrm{Clus}}=0.570$ \citep[][]{Postman:2012ca}. 

The \spitzer/IRAC imaging data in $3.6$ and $4.5$ $\micron$ are obtained from SURFS UP \citep{Bradac:2014el} supplemented with shallower data from The IRAC Lensing Survey (PI: Egami). The IRAC images reach a total integration time of $\sim 30$ hours in each band within the \HST field of view. We follow the same procedure as in \citet{Huang:2016} for IRAC photometry: the \HST positions and morphologies are used as the high-resolution prior in the template-fitting code \textsc{T-PHOT} \citep{Merlin:2015bw}, and IRAC PSFs are measured by stacking stellar objects found in both the main and flanking fields. 

A summary of the photometric properties is listed in Table \ref{tab:summary}. We note that in the F850LP image, Image A has a nominal $3.3\sigma$ detection in the isophotal aperture that we think is likely spurious based on morphology. Therefore, we exclude the F850LP flux density of Image A in our analysis, although including it does not change our results significantly. Image A is also severely blended with its neighbor in the IRAC images, so we can only assign conservative upper limits to its IRAC flux densities. 
Both Images B and C are in relatively clean regions that are free from photometric foreground contamination, although neither has significant detections in the IRAC images.

\section{Spectroscopic Constraints}\label{sec:spec}

\subsection{Keck DEIMOS Data}
We targeted MACS2129 with Keck DEIMOS on 2014 September 01, 2015 May 15--16, and 2015 October 16 (all dates are UT). All the exposures were taken under generally photometric conditions, with seeing $\lesssim1\arcsec$. The slitmasks with $1\arcsec$ wide slits were designed to include $z\gtrsim7$ LBG candidates and, when slits were available, to also include targets whose photometric redshifts had considerable probability at $z\geq6$. 

\begin{figure}[t]
\includegraphics[width=\columnwidth]{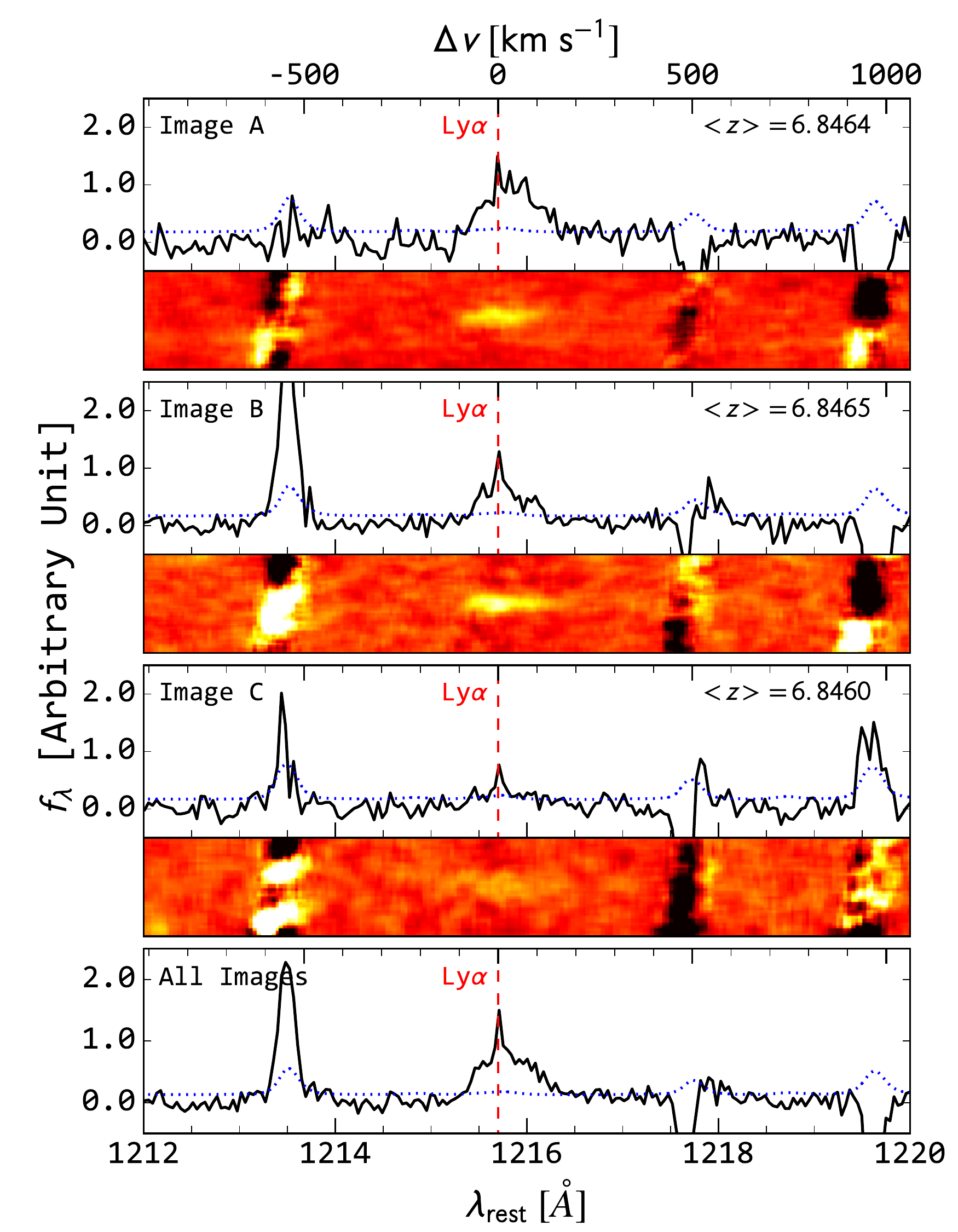}
\caption{Full-depth 1D and 2D Keck/DEIMOS spectra for each image. In the 1D spectra panels, flux density is shown in solid lines, and RMS is shown in dotted lines. The rest-frame wavelength scale is shown on the bottom and the velocity scale is shown on top. The stacked 2D spectra are smoothed by a boxcar filter of width 5 pixels. Spectra are shifted slightly to align the peak wavelengths (shown by a vertical dashed line), and the \lya redshifts determined by the peaks are indicated in the top-right corners.\label{fig:deimos}}
\end{figure}

The stacked 2D and 1D spectra from all observing runs are shown in Figure \ref{fig:deimos}. After stacking, we obtain significant line detections around 9538\AA: before correcting for slit loss, the measured line fluxes for Images A, B, and C are ($9.9\pm0.6)\times 10^{-18}$, ($5.7\pm0.6)\times10^{-18}$, and ($3.7\pm0.6)\times10^{-18}$ erg\,s$^{-1}$\,cm$^{-2}$, respectively. Even for Image C, we achieve a $6.1\sigma$ detection from the stacked 1D spectrum. We account for a slit throughput of 0.8 for a source with half-light radius of $0\farcs3$ under $0\farcs8$ seeing \citep{Lemaux:2009fy} with additional corrections made for conditions and bulk astrometric offsets. The slit loss-corrected line fluxes and \lya line luminosities are listed in Table \ref{tab:summary}. 

We measure the \lya equivalent widths (EWs) of each image using the F105W magnitudes to estimate the continuum level. The EWs of each image are reported in Table \ref{tab:summary}, and the weighted average EW (by S/N) is $74 \pm15$\AA{}. Because the S/N ratios are higher in DEIMOS data than in \emph{HST} grism data (Section \ref{subsec:grism}), we use the DEIMOS measurement as the fiducial values. The high EW, coupled with the blue rest-frame UV continuum slope $\beta=-2.76^{+1.23}_{-1.12}$ (measured from the stacked \HST images of the three sources), implies a high Ly$\alpha$ escape fraction of $\gtrsim$10\% from local Ly$\alpha$-emitting galaxies \citep{Hayes:2014jv,Henry:2015ht}. The Ly$\alpha$ escape fraction could also be crudely estimated by converting the Ly$\alpha$ luminosity into star formation rate (SFR) and compared with the UV-derived SFR. With this method, we estimate the Ly$\alpha$ escape fraction to be $\gtrsim$50\% assuming no dust correction for rest-frame UV flux. 

We fit a truncated Gaussian profile to the stacked 1D spectrum of all images to estimate the line width and find the half width at half maximum (HWHM) on the red side of the line to be $145\pm8~\mathrm{km}~\mathrm{s}^{-1}$ (random error only, after accounting for an instrumental resolution of 1.93\AA). The emission lines for Images A and B are individually broad enough to be resolved by the 1200G grating of DEIMOS, so we measure their line asymmetries $1/a_\lambda\equiv (\lambda_c-\lambda_{10,b})/(\lambda_{10,r}-\lambda_c)$, where $\lambda_c$ is the central wavelength of the emission, and $\lambda_{10,r}$ ($\lambda_{10,b}$) is the wavelength where the flux first exceeds 10\% of the peak redward (blueward) of the peak. The line asymmetries $1/a_\lambda$ for Images A and B are 0.35 and 0.71, consistent with the range of values for \lya emission showing depressed blue wings. Therefore, the line shapes also support the \lya interpretation.  

\begin{figure*}[t]
\includegraphics[width=\textwidth,trim={0 0.cm 0 0},clip]{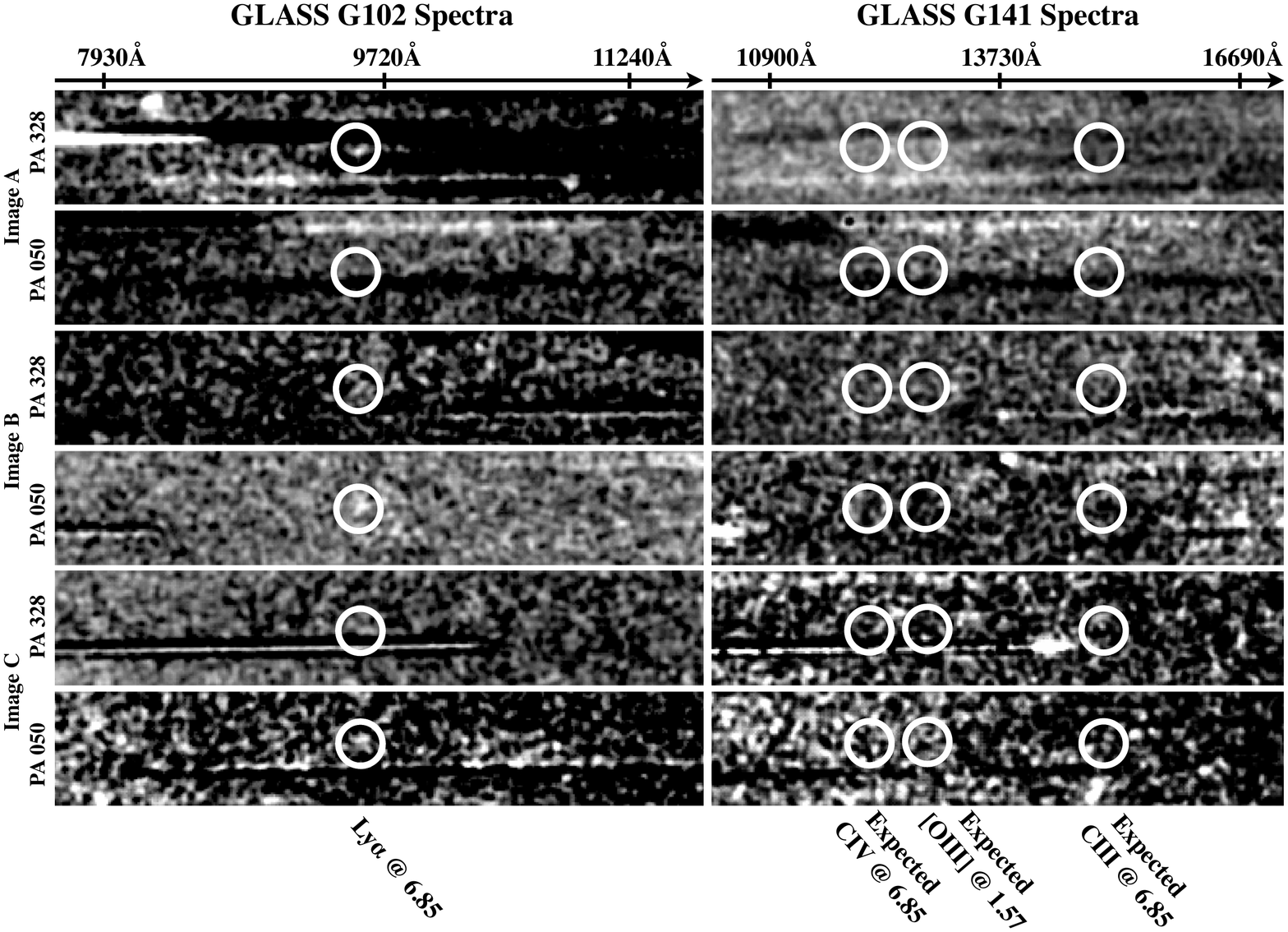}
\caption{Contamination-subtracted \HST G102 and G141 grism spectra for Images A, B, and C, obtained as part of the GLASS program. The grism data were taken at two different position angles (PAs; 50\degree and 328\degree), and we show each PA separately. The wavelengths of the \emph{observed} \lya and the \emph{expected} \civ, and \ciii lines at $z=6.85$ are marked by white circles.
We also mark the \emph{expected} locations of the \oiii line at $z=1.57$. We do not detect any significant flux at the expected wavelength of the potential \oiii line, and the upper limits on \oiii/\oii strongly support the $z=6.85$ \lya interpretation.\label{fig:grism}}
\end{figure*}

\subsection{\HST Grism Data}\label{subsec:grism}
\HST WFC3/IR G102 and G141 grism data for MACS2129 were taken as part of the Grism Lens-Amplified Survey from Space (GLASS) program \citep{Schmidt:2014kx,Treu:2015kr}. GLASS obtained 10-orbit G102 data and 4-orbit G141 data for each cluster, aiming to probe the \lya LF at $z\gtrsim 6$ and reaching uniform sensitivity across the two grisms that continuously cover $0.81\micron$ -- $1.69\micron$. For each cluster field, grism data were taken in two independent position angles (PAs; 50\degree and 328\degree for MACS2129) to facilitate contamination removal. 

\citet{Schmidt:2016ez} published the grism search on 159 LBGs from the first six clusters covered by GLASS, including MACS2129. They showed that 24 out of 159 LBGs have emission line detections consistent with being \lya, including Images B and C in this Letter; 
both sources have emission line detections around $9570$~\AA{}, corresponding to a \lya redshift $z=6.87$. However, due to the wavelength uncertainty of the grism spectra ($\sim30$~\AA), the \lya redshift cannot be determined to better than $\Delta~z=0.02$. We show the extracted grism spectra for all three images in Figure \ref{fig:grism} and the measured \lya fluxes in Table \ref{tab:summary}.

Image A is not in the $z\gtrsim7$ LBG sample of \citet{Schmidt:2016ez} because it was not selected as an LBG at $z \sim 6-7$ by multiple groups due to the spurious F850LP flux. But extraction of the grism spectra for Image A also shows an emission line at $\sim9570$~\AA{} ($\sim4\sigma$ detection). This is observed in the $\mathrm{PA}=328$\degree spectrum not affected by contamination, consistent with being \lya emission at the same redshift as Images B and C. 

Any (or all) of the three emission lines detected in grism spectra could also be the (unresolved) [O II] line at $z=1.57$, but our Keck DEIMOS spectra do not show any signs of resolved \oii doublets even though the spectral resolution is sufficient. Furthermore, the G141 grism data provide the wavelength coverage to detect (or rule out) \oiii emission lines at $z=1.57$. We do not find significant detections at the expected \oiii wavelength in the G141 spectra, and the $2\sigma$ upper limits for $f_{\mathrm{\oiii}}/f_{\mathrm{\oii}}$ range between 0.23 and 0.61, with a median $f_{\mathrm{\oiii}}/f_{\mathrm{\oii}}<0.35$. At $z=1.57$, this galaxy would have had a stellar mass of $\sim2\times10^8~\msolar$ and a sub-solar metallicity according to the mass-metallicity relation derived at $z\sim2$, which implies $f_{\mathrm{\oiii}}/f_{\mathrm{\oii}}\geq2$ \citep{Henry:2013gx}. Therefore, we rule out the \oii interpretation of the emission lines.

\begin{figure}[ht]
\includegraphics[width=\columnwidth,trim={2cm 0 2cm 2cm},clip]{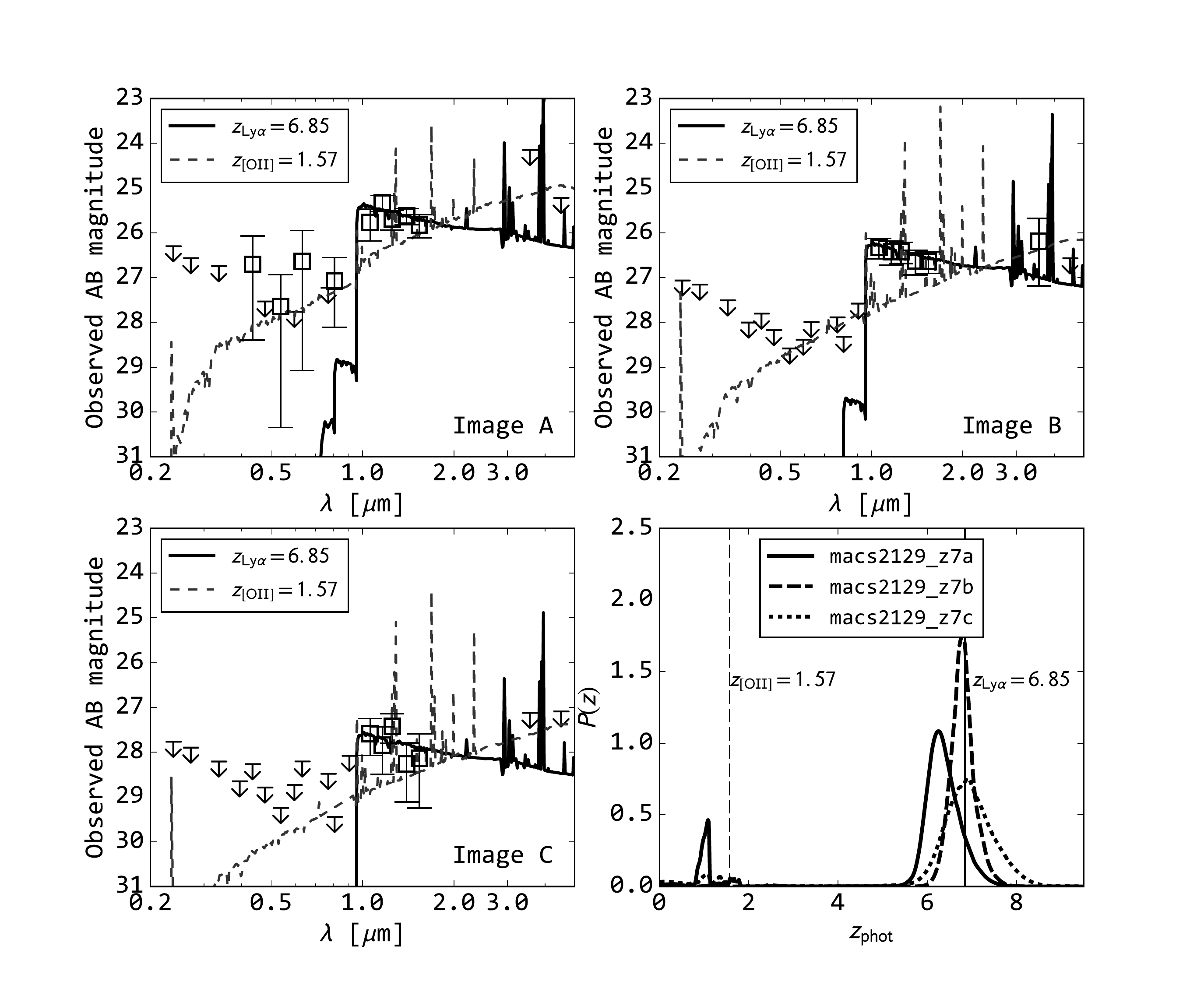}
\caption{Best-fit $0.2~Z_\odot$ SED templates for Images A, B, and C when the template redshift is fixed at $z_{\mathrm{Ly}\alpha}=6.85$ (solid line) and at $z_{\mathrm{\oii}}=1.57$ (dashed line). Images A, B, and C are shown in the upper left, upper right, and bottom left panel, respectively. The photometric redshift probability distributions are shown in the lower right panel. All three images are better fit by a $z=6.85$ template than a $z=1.57$ template.\label{fig:sedfit}}
\end{figure}

Other rest-frame UV nebular emission lines like \civ$\lambda$1549 and \ciii$\lambda$1909 have been proposed as alternative features for redshift confirmation at $z\geq6$ because they are not attenuated by the intergalactic medium \citep[e.g.,][]{Stark:2015if}. Detections of \civ and \ciii would also allow photoionization modeling to constrain the ionization parameter and metallicity \citep[e.g.][]{Erb:2010iy}.
%
%
%
%
The GLASS G141 spectra cover the wavelengths of both \civ and \ciii at $z=6.85$, but we do not detect any \civ or \ciii emission. The expected locations of \civ and \ciii in the G141 spectra are marked by white circles in Figure~\ref{fig:grism}, and we list the $1\sigma$ EW limits for \civ and \ciii in Table~\ref{tab:summary}. The most stringent limit on the rest-frame \civ and \ciii EW is $\lesssim25$\AA{} (measured from Image A with $\mu=11$), still not quite sufficient to detect the typical EWs of \civ and \ciii emission ($\leq25$\AA{}) from low-metallicity galaxies \citep[e.g.,][]{Stark:2014fa,Rigby:2015jy,Stark:2015if,Stark:2015dj}. 
\citet{Schmidt:2016ez} also derived a $2\sigma$ upper limit of \civ (\ciii) to \lya flux ratio of $0.32$ ($0.23$) from a stack of eight $z\gtrsim7$ \lya-emitting galaxies. Therefore, detecting the \civ and \ciii emission at $z\gtrsim7$ is still quite challenging from the available \HST grism spectra.
%

\section{Gravitational Lens Modeling}\label{sec:lensing}

Gravitational lens modeling of MACS2129 is made difficult by the sparsity of known multiple image systems with spectroscopic redshifts. \citet{Zitrin:2015gu} presented the first two lens models of the cluster using the CLASH photometry. 
While both models are constrained using 8 multiple image systems, only one of them \citep[system 1 in][]{Zitrin:2015gu} was spectroscopically confirmed \citep[see also][]{Christensen:2012dv}. Moreover, system 1 is a rare example of a sextuply-imaged galaxy with a spectroscopic redshift $z=1.364$; the unusual configuration and multiplicity of this system makes the modeling challenging, and neither model reproduces the multiplicity of system 1. The \citet{Zitrin:2015gu} models do reproduce the positions of the $z=6.85$ triple-image system accurately, but they predict that Image B is roughly twice as magnified as Image A, in disagreement with the observed ratio.


To improve the estimates of the absolute magnifications of the triply-imaged galaxy, we create a new lens model of MACS2129 using additional spectroscopic redshifts from GLASS and from CLASH-VLT (186.A-0798; PI: P. Rosati; Monna et al. in prep.). We identify a new $z=1.04\pm0.01$ multiple-image system that is not identified by \citet{Zitrin:2015gu}, and while we have spectroscopic redshifts for parts of other multiple-image systems, we are not confident enough in any other system as a whole. To be conservative, we model the cluster using system 1, the aforementioned new system at $z=1.04$, and the triply-imaged system at $z=6.85$. We employ the lens modeling method SWUnited \citep{Bradac:2005ex,Bradac:2009bk}, which constrains the gravitational potential of the cluster on a grid via an iterative $\chi^2$ minimization algorithm. We find that our lens model is able to reproduce the positions and relative magnifications of the three images at $z=6.85$, while also consistently fitting the other two systems used in the model. We therefore adopt the absolute magnifications of $11.0$, $5.4$, and $2.6$ for Images A, B, and C, respectively.

\section{Discussion}\label{sec:discussion}

We model the physical properties of the three images following the procedure outlined in \citet{Huang:2016}. In short, we adopt the \citet{Bruzual:2003ck} (BC03) templates with $0.2\,Z_\odot$ and a constant star formation history (SFH), motivated by the good recovery of SFRs and stellar masses of simulated galaxies \citep{Salmon:2015iz}. 
We also account for dust attenuation internal to the galaxy following the prescription in \citet{Calzetti:2000iy}, parameterized by $E(B-V)_s$ from 0 to 1. The templates also include strong nebular emission lines, whose fluxes are determined by the Lyman continuum flux of BC03 models and nebular line ratios from \citet{Anders:2003ci}. The fitting is done using the photometric redshift code EAZY \citep{Brammer:2008gn}. In Figure \ref{fig:sedfit} we show the best-fit $0.2\,Z_\odot$ templates at the \lya redshift $z_{\mathrm{Ly}\alpha}=6.85$ and at the \oii redshift $z_{\mathrm{\oii}}=1.57$. In the bottom right panel, we also show the photometric redshift probability distribution $P(z)$ for each image. All three images have $P(z)$ more consistent with the \lya redshift than with the \oii redshift based on the total $\chi^2$ values.

The modeling results are summarized in Table \ref{tab:summary}, where we report the best-fit values, 68\% confidence intervals (and 84 percentile upper limits where they apply). After accounting for lens magnification, we find that this galaxy has rest-frame 1600\AA{} absolute magnitude (converted from the observed F125W magnitudes) of $-18.6\pm0.2$ mag, stellar mass of $(1.5\pm0.3)\times10^7~\msolar$, SFR of $1.4\pm0.2~\sfrunit$, and specific SFR (sSFR) of $95\pm25~\ssfrunit$ (all are S/N-weighted averages). The galaxy is also best fit by a very young ($\lesssim 20~\mathrm{Myr}$ old), dust-free template. The rest-frame UV slope $\beta$---measured from the median-stacked images---is $-2.76^{+1.23}_{-1.12}$, although the errors are large due to low S/N in the WFC3/IR filters. The results suggest that this galaxy is one of the intrinsically faintest \lya-emitting galaxy confirmed at $z>6$ to date; its rest-frame UV luminosity is roughly $0.14~L_{\mathrm{UV},~z\sim 7}^*$ \citep[adopting $L_{\mathrm{UV},~z\sim 7}^*=-20.87 \pm 0.28$;][]{Bouwens:2015gm}, well into the faint end of the $z\sim7$ UV LF. With the detection of \lya emission from this galaxy (and from other similar galaxies), we start to probe the likely sources that dominated reionization: the low-mass, young, and (almost) dust-free galaxies \citep[e.g.,][]{Robertson:2015ia}.

Recently, several galaxies at $z\gtrsim7$ have been confirmed via their \lya emission (e.g., \citealt{Oesch:2016ik}, \citealt{2016arXiv160202160S}, and references therein).
Most of these galaxies are surprisingly bright, with $M_\mathrm{UV}$ ranging from $-20.5$ to $-22.4$ mag; 
some of the aforementioned galaxies are selected based on their unusual \emph{Spitzer}/IRAC colors, which imply very high \oiii or H$\alpha$ EWs. Due to the depths of the available \emph{Spitzer} imaging data, $z\gtrsim7$ galaxies identified this way are mostly the bright ones \citep[e.g.,][]{RobertsBorsani:2015um}.
It is possible that these bright galaxies are sitting in overdense regions inside local HII bubbles that enhance \lya transmission \citep[and references therein]{Dijkstra:2014iq} and offer a biased view of the reionization due to cosmic variance \citep{Trenti:2008hq}. On the other hand, fainter galaxies hosted by halos with mass $M\sim10^{9-10}~\msolar$ are more likely to dominate the ionizing photon budget, the details of which will depend on a few factors like gas cooling, supernovae feedback, and ionizing photon escape fraction \citep[e.g.,][]{2016arXiv160207711M}. So far all of the sub-$L^*$ galaxies at $z\gtrsim7$ confirmed via their \lya emissions are detected behind galaxy clusters (see also \citealt{Bradac:2012hi} and \citealt{Balestra:2013bj}); a larger sample of spectroscopically confirmed $z\gtrsim7$ galaxy will be valuable for understanding the reionization process, e.g., through the \lya fraction among LBGs \citep{Pentericci:2014cw}.

In several ways, the galaxy probed here is similar to the $z=6.4$ galaxy detected behind MACS0717$+$3745 \citep{Vanzella:2014cm}. Both galaxies have faint UV luminosities ($\lesssim0.2~L^*$), low stellar masses ($\lesssim5\times10^7~\msolar$), \lya HWHM of $\sim100$--$150$ km s$^{-1}$, and possibly high \lya escape fractions ($\geq10$\%). 
Perhaps the best local analogs of both galaxies are the Green Pea galaxies \citep[e.g.,][]{Henry:2015ht}, which have high \lya EWs ($10$--$160$\AA), low stellar masses ($10^8$--$10^9~\msolar$), and blue UV slopes ($\beta\sim-2.2$ to $-1.6$) that suggest very low dust content. The detections of these sub-$L^*$ galaxies at $z\gtrsim7$ are made possible due to gravitational lensing, which will be the most efficient way to detect galaxies even in the \emph{JWST} era.
 
\vspace{2em}
We thank the referee for constructive feedback of this work. We also thank Piero Rosati and Anna Monna for providing spectroscopic redshifts from the CLASH-VLT program. This work is based on observations made with the NASA/ESA Hubble Space Telescope, obtained at the Space Telescope Science Institute, which is operated by the Association of Universities for Research in Astronomy, Inc., under NASA contract NAS5-26555 and NNX08AD79G and ESO-VLT telescopes. Observations were carried out using Spitzer Space Telescope, which is operated by the Jet Propulsion Laboratory, California Institute of Technology under a contract with NASA. Support for this work is provided by NASA through a Spitzer award issued by JPL/Caltech, HST-AR-13235, HST-GO-13459, and HST-GO-13177. AH acknowledges support by NASA Headquarters under the NASA Earth and Space Science Fellowship Program Grant ASTRO14F-0007.


\end{document}